\documentclass[10pt,twocolumn,preprintnumbers,amsmath,amssymb,nofootinbib]{revtex4-2}
\usepackage{amsmath,amssymb,amsthm,mathrsfs}
\usepackage{epsfig}
\usepackage{graphicx}
\usepackage[usenames,dvipsnames]{color}
\usepackage{subfigure}
\usepackage{slashed}
\usepackage{comment}
\usepackage[normalem]{ulem}
\usepackage{url}
\usepackage[colorlinks, citecolor=blue,urlcolor=blue,linkcolor=blue,runcolor=filecolor]{hyperref}

\newcommand{\chih}{\hat{\chi}}

\begin{document}
\title{Frame-dragging effects in a gravitational quantum field theory}

\author{Dongfeng Gao\textsuperscript{1,2,}}
\altaffiliation{Email: dfgao@wipm.ac.cn}
\author{Wei-Tou Ni\textsuperscript{1,3,}}
\altaffiliation{Email: wei-tou.ni@wipm.ac.cn}

\vskip 0.5cm
\affiliation{1 State Key Laboratory of Magnetic Resonance and Atomic and Molecular Physics, Wuhan Institute of Physics and Mathematics, Innovation Academy for Precision Measurement Science and Technology, Chinese Academy of Sciences, Wuhan 430071, China\\
2 Hefei National Laboratory, Hefei 230088, China\\
3 International Centre for Theoretical Physics Asia-Pacific, University of Chinese Academy of Sciences, Beijing 100190, China}

\date{\today}
\begin{abstract}
\noindent 

Analogous to magnetism in electrodynamics, it is gravitomagnetism in relativistic gravity.  Since gravity determines locally inertial frames, in general relativity (GR) and other relativistic theories of gravity, frame-dragging with source motion plays a key role in gravitomagnetism. Recently, Wu has put forward a gauge theory of gravity, called the gravitational quantum field theory (GQFT), with the gravitational force and the spin gauge force described by the gauge fields. Gao {\it et al.} ({\it Phy. Rev. D 109, 064072}) have derived the Shapiro time delay in the GQFT and given an empirical constraint from the Cassini experimental result on the dimensionless GQFT parameter $\gamma_W$ to be $(2.1\pm 2.3)\times 10^{-5}$. In this work, we derive the frame-dragging Lense-Thirring effects in the GQFT. The current precision of LARES-LAGEOS Lense-Thirring measurement gives a constraint on $|\gamma_W|$ to be less than $2\times 10^{-2}$. This constraint is consistent with, but subdominant to, the Cassini experimental constraint. As a candidate of quantum gravity, we do not expect that the deviation from the GR value  ($\gamma_W=0$) is large, classically. With the launch of LARES 2, the precision of the Lense-Thirring measurement is expected to increase by one order of magnitude in a couple of years. As to the Shapiro effect, current technologies have the capability to measure the $\gamma_W$ parameter to a precision of $10^{-9}$.

\end{abstract}

\maketitle
%\newpage

\section{Introduction}\label{s1}

The classical theory of gravitation is Einstein’s theory of general relativity (GR), which has passed various tests from both astrophysical observations and experimental measurements \cite{Will:2014kxa, Ni2015}. In the genesis of GR, Einstein was originally motivated by Mach’s principle. However, in his formulation, GR is firmly founded on the equivalence principle and local special relativistic physics. The key to reconcile these two concepts is that for moving/rotating bodies, the locally inertial frames (Minkowski physics) around them are dragged by the motion/rotation of them. Lense and Thirring \cite{Thirring1918a,Thirring1918b, Mashhoon1984} discovered this effect in communication with Einstein. Experimentally, this effect was discovered in 2004 by Ciufolini and Pavlis \cite{Ciufolini2004}, and in 2011 by Everitt {\it et al.} \cite{PhysRevLett.106.221101}.

In GR, the gravity is solely described by the geometric property of the spacetime. As comparison, the other three fundamental interactions (the electromagnetic, weak, and strong interactions) of nature are all described by quantum field theories, forming successfully into the standard model of particle physics in the end \cite{RevModPhys.71.S96}. Following the similar spirit, there are many studies tempting to describe the gravity in terms of the gauge theory language (see Refs.~\cite{Hehl:1976kj,Ivanenko:1983fts,Hehl:1994ue} for review).

Recently, Wu put forward a gauge theory of gravity, called the gravitational quantum field theory (GQFT)~\cite{Wu:2022aet,Wu:2022mzr}. In the model, the gravitational force and the spin gauge force are described by the gauge fields corresponding to the inhomogeneous spin gauge symmetry, ${\rm WS}(1,3) = {\rm SP}(1,3)\rtimes {\rm W}^{1,3}$. (${\rm SP}(1,3)\equiv {\rm SO}(1,3)$ is the spin gauge symmetry group, ${\rm W}^{1,3}$ is the translation-like chiral-type gauge symmetry group, and $\rtimes$ means a semi-direct product.). Gao {\it et al.} \cite{PhysRevD.109.064072} derived the Shapiro time delay in this theory. It would be interesting to derive the frame-dragging Lense-Thirring effects in this theory. We do so in this paper.

The paper is organized as follows. In Sec.~\ref{themodel}, we will introduce the linearized GQFT in Refs. \cite{Wu:2022mzr,PhysRevD.109.064072}. In Sec. \ref{solutionGQFT}, we will discuss the linearized GQFT under the harmonic gauge condition. In Sec. \ref{fieldeq}, we derive in detail the slow motion metric with sources. Especially, we derive the frame-dragging effects in the GQFT. In Sec. \ref{lensethirring}, we discuss the current experimental prospect on the measurement of Lense-Thirring effects. Finally, the conclusion and outlook are given in Sec. \ref{conclusion}.

\section{the linearized gravitational equations of the GQFT}\label{themodel}
In this section, we will summarize the derivation of the linearized gravitational equations in Ref. \cite{PhysRevD.109.064072}. 

%\subsection{The linearized GQFT}
Let us start with the gauge-type gravitational equation (Eq. (159) of Ref. \cite{Wu:2022mzr}),
\begin{equation}  \label{GTGE}
\partial_{\nu} \tilde{F}^{\mu\nu }_{a}  = J_{a}^{\; \mu}  , 
\end{equation}
where the field strength $\tilde{F}^{\mu\nu }_{a}$ is defined by the following:
\begin{eqnarray}
\tilde{F}^{\mu\nu }_{a}  &\equiv&  \chi \, \tilde{\chi}^{[\mu\nu]\mu'\nu'}_{a a'}  F_{\mu'\nu' }^{a'}\, ,\nonumber \\
F_{\mu'\nu'}^{a'}(x) &\equiv& \partial_{\mu'}\chi_{\nu'}^{\; a'}(x) -  \partial_{\nu'}\chi_{\mu'}^{\; a'}(x) \, ,\nonumber \\
\tilde{\chi}_{aa'}^{[\mu\nu] \mu'\nu'} &\equiv& \chih_{c}^{\;\, \mu}\chih_{d}^{\;\, \nu} \chih_{c'}^{\;\, \mu'} \chih_{d'}^{\;\, \nu'}  \tilde{\eta}^{[c d] c' d'}_{a a'} \, ,\nonumber \\
\tilde{\eta}^{[c d] c' d'}_{a a'} &\equiv& \frac{1}{2}(  \tilde{\eta}^{cd c' d'}_{a a'}  - \tilde{\eta}^{ dc c' d'}_{a a'} )\, ,\nonumber \\
\tilde{\eta}^{cd c'd'}_{a a'}  & \equiv &    \eta^{c c'} \eta^{d d'} \eta_{a a'}  
+  \eta^{c c'} ( \eta_{a'}^{d} \eta_{a}^{d'}  -  2\eta_{a}^{d} \eta_{a'}^{d'}  ) \nonumber \\
&&+  \eta^{d d'} ( \eta_{a'}^{c} \eta_{a}^{c'} -2 \eta_{a}^{c} \eta_{a'}^{c'} )  \, , \nonumber \\
\chi &\equiv& {\rm det} \, \chi_{\mu}^{\; a}\, .
\end{eqnarray}
The source $J_{a}^{\; \mu}$ contains various contributions, as defined in Ref. \cite{Wu:2022mzr}. One can see that the basic ingredient in the gravidynamics is the gravigauge field $\chi_{\mu}^{\, a}$ (tetrad). To linearize the GQFT, we decompose it into 
\begin{eqnarray}\label{Gpert}
	\chi_{\mu}^{\,a} \equiv \eta_{\mu}^{\,a} + \frac{1}{2} h_{\mu}^{\, a}\,,
\end{eqnarray} 
where $\eta_{\mu}^{\, a}$ is regarded as the background field, and $h_{\mu}^{\, a}$ is assumed to be a weak perturbation. Then, the dual gravigauge field can be written as
\begin{equation} 
\hat{\chi}^{\,\mu}_a = (\eta_{\mu}^{\,a} + \frac{1}{2} h_{\mu}^{\, a} )^{-1}=\eta^{\,\mu}_{a} - \frac{1}{2} h^{\,\mu}_{a}\, .
\end{equation}

After a bit of calculation, one can find that, to the leading order in $h^{\,\mu}_{a}$, 
\begin{equation*} 
\partial_{\nu} \tilde{F}^{\rho\nu }_{a}=\square h_a^{\,\rho} - \partial^\rho \partial_\nu h_a^{\,\nu} - \partial_\nu \partial_a h^{\nu \rho} + \partial_a \partial^\rho h + \eta^{\,\rho}_a (\partial_\nu\partial_\sigma h^{\nu \sigma} -\square h )
\end{equation*}
where $\square\equiv \partial_\mu\partial^\mu={\rm \partial_0^{\, 2}}+{\rm \partial_i}{\rm \partial^i}$.
Do the similar calculation for the terms involving $h^{\,\mu}_{a}$ in the source $J_{a}^{\; \mu}$, one finally gets the linearized gravitational equations
\begin{widetext}
	\begin{eqnarray}
		&&	\left[ \square h_a^{\,\rho} - \partial^\rho \partial_\nu h_a^{\,\nu} - \partial_\nu \partial_a h^{\nu \rho} + \partial_a \partial^\rho h + \delta^\rho_a (\partial_\nu\partial_\sigma h^{\nu \sigma} -\square h )  \right] + \gamma_W (\square h_a^{\,\rho}-\partial^\rho \partial_\nu h_a^{\,\nu})  = -16 \pi {G}_\kappa {\rm J}_a^{\,\rho}\,,
	\end{eqnarray} 
\end{widetext}
where the $\gamma_W$-dependent terms come from the relevant part in $J_a^{\,\mu}$, and $\gamma_W \equiv \gamma_G (\alpha_G- \alpha_W/2)$ defined in Ref. \cite{Wu:2022mzr}. Also, $h_a^{\,\mu}$ has been gauge fixed to be symmetric, $h_a^{\,\mu} = h^\mu_{\,\,a}$. One also has $h\equiv \eta_a^{\,\mu} h^a_{\,\mu}$, and $h_{\mu\nu}\equiv \eta^a_{\,\nu} \eta_{\mu\rho} h^{\,\rho}_{a}$.

Furthermore, one can decompose the above equation into the symmetric and anti-symmetric parts as the following:
\begin{widetext}
	\begin{eqnarray}\label{eomGTsym}
		\widetilde{G}_{(\mu\nu)}\equiv \frac{1}{2}\left[\square h_{\mu\nu} - 2\partial^\sigma \partial_{(\mu} h_{\nu)\sigma} + \partial_\mu\partial_\nu h + \eta_{\mu\nu} (\partial^\rho\partial^\sigma h_{\rho\sigma} -\square h ) \right] + \frac{\gamma_W}{2} [\square h_{\mu\nu} - \partial^\sigma \partial_{(\mu} h_{\nu)\sigma}]
		= -8\pi G_\kappa T_{(\mu\nu)}\,, 
	\end{eqnarray} 
\end{widetext}
\begin{eqnarray}\label{eomGTanti}
	\widetilde{G}_{[\mu\nu]} \equiv -\frac{\gamma_W}{2} \partial^\sigma \partial_{[\mu} h_{\nu]\sigma} = -8\pi {G}_\kappa T_{[\mu\nu]}\,,
\end{eqnarray} 
where $T_{(\mu\nu)}$ and $T_{[\mu\nu]}$ stand for the symmetric and anti-symmetric parts of the source. We adopt the convention that $\eta_{\mu\nu}\equiv {\rm diag}(1,-1,-1,-1)$ and  $h=h_{00}-h_{11}-h_{22}-h_{33}\equiv h_{00}-h_{ii}$. Now, it is clear that Eq. (\ref{eomGTsym}) can be regarded as a generalized Einstein's equation. Eq. (\ref{eomGTanti}) has no counterpart in GR, and plays important roles in the linearized GQFT.

\section{The linearized GQFT under the harmonic gauge condition} \label{solutionGQFT}

To solve the linearized GQFT, we should apply some type of gauge conditions to Eqs. (\ref{eomGTsym}) and (\ref{eomGTanti}). Following Ref. \cite{Ni2015}, we adopt the harmonic gauge condition for $h_{\mu\nu}$,
\begin{eqnarray}\label{gaugeh}
\partial^{\nu} h_{\mu\nu}= \frac{1}{2}\partial_{\mu} h\, .
\end{eqnarray} 
It is easy to check that Eq.~\eqref{eomGTanti} is consistent with the gauge condition (\ref{gaugeh}) if $T_{[\mu\nu]}=0$.  In the following, we will assume $T_{[\mu\nu]}=0$, as done in Ref. \cite{PhysRevD.109.064072}. 

With Eq. (\ref{gaugeh}), the linearized GQFT equation (\ref{eomGTsym}) becomes
\begin{eqnarray}\label{eomGTsym1}
(1+\gamma_W)\square h_{\mu\nu} -&& \frac{1}{2}(\eta_{\mu\nu}\square+ \gamma_W \partial_\mu \partial_{\nu}) h \nonumber\\
     &&= -16\pi G_\kappa T_{(\mu\nu)}\, .
\end{eqnarray} 

By contracting $\eta^{\mu\nu}$ with Eq. (\ref{eomGTsym1}), one can find that
\begin{eqnarray*}\label{solutionhh}
	\square h= \frac{16\pi G_\kappa}{1-\gamma_W/2} T\, ,
\end{eqnarray*} 
where $T\equiv {\rm Tr}(T_{(\mu\nu)})$.
Inserting it into Eq. (\ref{eomGTsym1}), we get
\begin{eqnarray}\label{eomGTsym2}
&&	(1+\gamma_W)\square h_{\mu\nu}-\frac{\gamma_W}{2} \partial_\mu \partial_{\nu} h\nonumber\\
&&=16\pi G_\kappa \left[\frac{\eta_{\mu\nu}}{2-\gamma_W} T-T_{(\mu\nu)}\right]\, .
\end{eqnarray} 

To be explicit, Eq. (\ref{eomGTsym1}) can be decomposed into the following components: 
\begin{itemize}
	\item the $(0,0)$ component
	\begin{eqnarray}
		\label{EQtt}
		[(\frac{1}{2}+\gamma_W) \square-\frac{\gamma_W}{2}\partial_0^2]&& h_{00}+\frac{1}{2}(\square+\gamma_W\partial_0^2) h_{ii} \nonumber\\
		&& = -16\pi {G}_\kappa T_{(00)}({\bf x})\, ,
	\end{eqnarray}
	\item the $(0,j)$ component
	\begin{eqnarray}
		\label{EQti}
		&&(1+\gamma_W)	\square h_{0j}-\frac{\gamma_W}{2} \partial_0 \partial_j (h_{00}-h_{ii})= -16\pi G_\kappa T_{(0j)} ,\,\,\,\,\,\,\,\,\,
	\end{eqnarray}
	\item the $(i,j)$ component
	\begin{eqnarray}	\label{EQij}
		(1+\gamma_W)\square h_{ij} - &&\frac{1}{2}(\eta_{ij}\square+ \gamma_W \partial_i \partial_j) (h_{00}-h_{kk})\nonumber\\
		&&	= -16\pi G_\kappa T_{(ij)}\, .
	\end{eqnarray} 
\end{itemize}
By contracting $\eta^{ij}$ with Eq. (\ref{EQij}), one can get a useful equation,
\begin{eqnarray}	\label{EQij'}
	(3\square+ \gamma_W \partial_k \partial^k) h_{00}+&&[(2\gamma_W-1)\square- \gamma_W \partial_k \partial^k]h_{ii}\nonumber\\
	&&	= -32\pi G_\kappa T_{(ii)}\, .
\end{eqnarray}

\section{Slow motion metric for the linearized equations with sources}\label{fieldeq}

Let us apply the slow motion approximation to the generalized Einstein equation, Eq. (\ref{eomGTsym1}). According to Refs. \cite{Chandrasekhar1965ThePE,Will1993}, in the slow motion limit, the following approximations are made for the metric: $h_{00}\sim \mathcal{O}(\epsilon^2)$, $h_{0i}\sim \mathcal{O}(\epsilon^3)$, and $h_{ij}\sim \mathcal{O}(\epsilon^2)$. For the source, one has $T_{(00)}\sim \mathcal{O}(\epsilon^0)$, $T_{(0i)}\sim \mathcal{O}(\epsilon^1)$, and $T_{(ij)}\sim \mathcal{O}(\epsilon^2)$, where $\epsilon= v/c$ denotes the order of smallness.

By keeping the lowest order terms in $\epsilon$, the generalized Einstein equations, Eqs. (\ref{EQtt})-(\ref{EQij'}), can be written as follows,
\begin{eqnarray}
\label{EQtt1} \partial_k\partial^k[(1/2+\gamma_W) h_{00}+ h_{ii}/2]  = -16\pi {G}_\kappa T_{(00)}\, ,
\end{eqnarray}
\begin{equation}
\label{EQti1} (1+\gamma_W)\partial_k\partial^k h_{0j}-\frac{\gamma_W}{2} \partial_0 \partial_j (h_{00}-h_{ii})= -16\pi G_\kappa T_{(0j)} ,\,\,\,\,\,\,\,\,\, \,\,
\end{equation}
\begin{eqnarray}	\label{EQij''}
(1+\gamma_W)\partial_k\partial^k h_{ij} = \frac{1}{2}(\eta_{ij}\partial_k\partial^k+ \gamma_W \partial_i \partial_j) h\, ,\,\,\,\,\,\,\,\,\,\,
\end{eqnarray} 
\begin{eqnarray}
\label{EQij1} (3+\gamma_W) h_{00}+(\gamma_W-1) h_{ii}  = 0\, .
\end{eqnarray}
Substituting Eq. (\ref{EQij1}) into Eq. (\ref{EQtt1}), one has
\begin{eqnarray}\label{solutiontt'}
\partial_k \partial^k h_{00}&&= - \frac{8 (1-\gamma_W)\pi {G}_\kappa}{(1-\gamma_W/2)(1+\gamma_W)} T_{(00)}\, \nonumber\\
&&\equiv -8 \pi G_N  T_{(00)}\, .
\end{eqnarray}
Thus, we reproduce the same relation between the Newtonian constant and the coupling ${G}_\kappa$ as in Ref. \cite{PhysRevD.109.064072},
\begin{eqnarray}\label{constG}
G_N= - \frac{1-\gamma_W}{(1-\gamma_W/2)(1+\gamma_W)} {G}_\kappa \, .
\end{eqnarray}
The solution to Eq. (\ref{solutiontt'}) is
\begin{eqnarray}\label{solutiontt}
h_{00}({\bf x}) =-2 {G}_N \int \frac{T_{(00)}({\bf x}')}{|{\bf x}-{\bf x}'|} d^3x'\, .
\end{eqnarray}
With Eqs. (\ref{EQij1}) and (\ref{solutiontt}), the solution to Eq. (\ref{EQij''}) is easy to find out,
\begin{eqnarray}\label{solutionij}
h_{ij}({\bf x})&=& -\eta_{ij} h_{00}/(1-\gamma_W)+h_{ij}^{(1)}\nonumber\\
&=&\frac{2 {G}_N\eta_{ij}}{1-\gamma_W} \int \frac{T_{(00)}({\bf x}')}{|{\bf x}-{\bf x}'|}d^3x'+h_{ij}^{(1)}\, ,
\end{eqnarray}
where
\begin{eqnarray*}
	h_{ij}^{(1)}({\bf x})&=& \frac{-\gamma_W}{4\pi(1-\gamma_W)} \int \frac{\partial_i\partial_j h_{00}({\bf x'})}{|{\bf x}-{\bf x}'|}d^3x'\, .
\end{eqnarray*}
	
Next, let us work out the solution to $h_{0j}$. Substituting Eq. (\ref{EQij1}) into Eq. (\ref{EQti1}), we have
\begin{eqnarray}	
\label{solutionti}  &&\partial_k\partial^k h_{0j}-\frac{\gamma_W }{(1-\gamma_W)} \partial_0 \partial_j h_{00}= - \frac{16\pi {G}_\kappa}{1+\gamma_W} T_{(0j)}\, .
\end{eqnarray}	

To solve Eq. (\ref{solutionti}), we need both the anti-symmetric gravitational equations \eqref{eomGTanti} and the gauge condition (\ref{gaugeh}). The (i,j) component of Eq. \eqref{eomGTanti} is 
\begin{eqnarray}
	\label{ANTIij1} \partial^0\partial_i h_{j0}+\partial^k \partial_i h_{jk}=\partial^0 \partial_j h_{i0}+\partial^k \partial_j h_{ik}\, .
\end{eqnarray}	
Under the slow motion limit, Eq. (\ref{gaugeh}) can be decomposed into 
\begin{eqnarray}
	\label{gaugeh0} 	\partial^{j} h_{0j}&=& -\frac{1}{2}\partial_0 (h_{00}+h_{ii})\, ,\\
	\label{gaugehi} 	\partial^{k} h_{jk}&=& \frac{1}{2}\partial_j (h_{00}-h_{ii})\, .
\end{eqnarray} 
From Eq. (\ref{gaugehi}), one can check that $\partial_i\partial^{k} h_{jk}=\partial_j\partial^{k} h_{ik}$. Applying this relation to Eq. (\ref{ANTIij1}), we find that 
\begin{eqnarray}
	\label{ANTIij2} \partial_i h_{j0}=\partial_j h_{i0}\, .
\end{eqnarray}
Eq. (\ref{ANTIij2}) means that
\begin{eqnarray}
	\label{ANTIij3} \partial_i \partial^i h_{j0}=\partial_j \partial^i h_{i0}\, .
\end{eqnarray}
Substituting Eq. (\ref{EQij1}) into  Eq. (\ref{gaugeh0}), we get
\begin{eqnarray}
	\label{gaugeh0'} &&	\partial^{k} h_{0k}=-\frac{2}{1-\gamma_W}\partial_0 h_{00}\, .
\end{eqnarray}	

Inserting Eqs. (\ref{ANTIij3}) and (\ref{gaugeh0'}) into Eq. (\ref{solutionti}), we finally get 
\begin{eqnarray}	
	\label{solutionti1}  &&\partial_k\partial^k h_{0j}= - \frac{16\pi {G}_\kappa} {(1-\gamma_W/2)(1+\gamma_W)} T_{(0j)}\, .
\end{eqnarray} 
With the relation (\ref{constG}), we can find out the solution to $h_{0j}$, which is
\begin{eqnarray}\label{solutionti2} 
	h_{0j}({\bf x})=\frac{-4 {G}_N}{1-\gamma_W} \int \frac{T_{(0j)}({\bf x}')}{|{\bf x}-{\bf x}'|} d^3x'\, .
\end{eqnarray}	
Note that it seems we have two different equations for $h_{0j}$, Eqs. (\ref{gaugeh0'}) and (\ref{solutionti1}).  In fact,  let us apply the 0-component of the conservation law, $\partial^{\nu}T_{(\mu\nu)}=0$, to Eq. (\ref{gaugeh0'}). With help of Eq. (\ref{solutiontt}), it is easy to check that Eq. (\ref{gaugeh0'}) reproduces exactly the same solution, Eq. (\ref{solutionti2}). \footnote{As a consistency check, one can check that the solutions (\ref{solutiontt}), (\ref{solutionij}), and (\ref{solutionti2}) of $h_{\mu\nu}$ do satisfy the gauge condition (\ref{gaugeh}). It is also straightforward to check that $h_{\mu\nu} \rightarrow h'_{\mu\nu} =h_{\mu\nu}+\mathcal{O}(h^2)$ under the general coordinate transformation, $x^{\mu} \rightarrow x'^{\mu} = x^{\mu} +\xi^{\mu}  \,\, {\rm with}\,\,\, \xi^{\mu}=\mathcal{O}(h)$.}

Comparing Eq. (\ref{solutionti2}) with the GR result, we finally get
\begin{equation}
	\label{EQti3} 
	h_{0j}=\frac{1}{1-\gamma_W} h_{0j}^{(GR)}\, .
\end{equation}
For a rotating system, the integral in Eq. (\ref{solutionti2}) can be evaluated to be
\begin{eqnarray}
\int \frac{T_{(0j)}({\bf x}')}{|{\bf x}-{\bf x}'|} d^3x'=\frac{1}{2}\frac{({\bf J}\times {\bf r})_j}{r^3}\, ,
\end{eqnarray}
where ${\bf J}$ is the angular momentum of the rotating system (e. g., see Refs. \cite{Weinberg1972, MTW1973,Ciufolini1995}).

Summing up this section, we obtain the linearized metric solution for the slow-motion weak-field rotating system:
\begin{eqnarray}
	g_{00}&=&1+2 G_N U/c^2 + \mathcal{O}(h^2, \epsilon^4)\, ,\\
\label{slowmetric} 	g_{0i}&=& h_{0i}=-\frac{2 G_N}{c^3(1-\gamma_W)}\frac{({\bf J}\times {\bf r})_i}{r^3} + \mathcal{O}(h^2, \epsilon^5)\, ,\\
	g_{ij}&=& -(1-2 G_N U/(c^2(1-\gamma_W)))\delta_{ij} + \mathcal{O}(h^2, \epsilon^4)\, ,
\end{eqnarray} 
where $U=-M/r$ is the gravitational potential of the rotating system. Note that it is easy to show that the leading term of $g_{0i}$ is of $\mathcal{O}(\epsilon^3)$.

The Lense-Thirring frame dragging effect comes from $g_{0i}$. The Shapiro time delay and the light deflection effect come from the second term in $g_{00}$ and the second term in $g_{ij}$. In the next section, we discuss the current status of the experimental measurements of frame-dragging effects and their constraint on the GQFT together with the current and prospects of the accuracy of determining the parameter $\gamma_W$.

\section{The Lense-Thirring tests of GQFT}\label{lensethirring}

When an electric charge is not in motion, it produces an electric field; when it moves, it produces a magnetic field. When matter is not in motion, it produces a gravitational (gravitoelectric) field; when it moves, it produces a gravitomagnetic field $g_{0i}$. From the nonvanishing of $g_{0i}$ at a point near the gravitating source, the local inertial (Lorentz frame) is rotating with respect to the coordinate frame with a rotating rate of $(G_N/c^3) {\bf J}\times {\bf r}/r^3$ ($\sim (G_N U/c^2) \omega $, with $\omega$ a typical/average rotating rate of the system). It is the gravitomagnetic field $g_{0i}$ that drags local reference frames with/around it with an attenuation factor of dimensionless gravity strength $G_N U/c^2$. This is an important aspect of relativistic gravity and a manifestation of gravitomagnetism. 

After Lense and Thirring published their work \cite{Thirring1918a,Thirring1918b, Mashhoon1984}, more than 100 years of theoretical research and experimental efforts have resulted in the astrodynamical measurements of the Gravity Probe B and the LAGEOS-LARES missions, achieving a precision of Earth's gravitomagnetic field at the level of 2\% \cite{Ciufolini2004,PhysRevLett.106.221101,Ciufolini2019}. With the launching of LARES 2 satellite in 2022, it is anticipated that one order of magnitude improvement can be achieved in a couple of years \cite{Ciufolini2023}. These results and anticipation are listed in Table \ref{table1}.

\begin{widetext}
	\begin{center}
		\begin{table}
			\caption{Current status of Lense-Thirring/Schiff effect measurements and the determination of $\gamma_W$}
			\label{table1}
			\begin{tabular}{|c|c|c|}		
				\hline
				% after \\: \hline or \cline{col1-col2} \cline{col3-col4} ...
				Frame-dragging experiment  & Accuracy in terms of GR& Type of experiment \\ 
				&  Lense-Thirring/Schiff  effect&  \\ \hline
				LARES-LAGEOS \cite{Ciufolini2004} & 10-30 \% &  Laser ranging orbit \\
				Gravity Probe B  \cite{PhysRevLett.106.221101}& 19 \% &  Gyros in orbit\\
				LARES-LAGEOS-LAGEOS 2  \cite{Ciufolini2019}& 2 \% &  Laser ranging orbit \\
				LARES-LARES 2-LAGEOS-LAGEOS 2 \cite{Ciufolini2023} & $\sim$ 0.2 \% (goal) &  Laser ranging orbit \\
				\hline
			\end{tabular}
		\end{table}			
	\end{center}	
\end{widetext}

\section{Conclusion and Outlook}\label{conclusion}

We have obtained the gravitomagnetic part of the metric and the Lense-Thirring/Schiff effects in the GQFT. The present satellite experiments have measured these effects for the Earth and constrain the absolute value of the the dimensionless GQFT parameter $\gamma_W$ to be less than $2\times 10^{-2}$ and an improvement of accuracy of one order is on the way. We do expect to be able to measure the frame dragging effect of the Sun and even possibly the Galaxy \cite{Ni2024}. But these experiments would be more a measurement of the solar and Galactic angular momenta. 

As a candidate for quantum gravity, we do expect the deviation of GQFT from GR to be small. In a previous investigation \cite{PhysRevD.109.064072}, the $\gamma_W$ parameter of GQFT is constrained to be $(2.1\pm 2.3)\times 10^{-5}$ by the Cassini-Shapiro time delay experiment. 

As to the ongoing experiments, Gaia Data Release 3 (Gaia DR3) \cite{Gaia} has already become public and contained astrometric results for more than 1 billion stars brighter than magnitude 20.7 based on observations collected by the Gaia satellite during the first 34 months of its operational phase. With the expected 4-year observation period, a simulation shows that GAIA could measure $\gamma$ to $1\times 10^{-5}$-$2\times 10^{-7}$ accuracy \cite{Gaia2016,Gaia2003}. This is listed as the second row in Table \ref{table2}. The analysis is still ongoing \cite{Gaia2023}.

BepiColombo mission \cite{Bepi} comprises two spacecrafts: the Mercury Planetary Orbiter (MPO) and the Mercury Magnetospheric Orbiter (MMO). Launched on 20 October 2018, arriving at Mercury in late 2025, it will gather data during its 1-year nominal mission, with a possible 1-year extension. Milani, Vokrouhlicky, Villani, Bonanno and Rossi \cite{PhysRevD.66.082001} have simulated the radio science of this mission: “While determining its orbit around Mercury, it will be possible to indirectly observe the motion of its center of mass, with an accuracy several orders of magnitude better than what is possible by radar ranging to the planet’s surface. This is an opportunity to conduct a relativity experiment which will be a modern version of the traditional tests of general relativity, based upon Mercury’s perihelion advance and the relativistic light propagation near the Sun.” They predict that the determination of $\gamma$ can reach $2\times 10^{-6}$.

ASTROD (Astrodynamical Space Test of Relativity using Optical Devices) I is envisaged as the first in a series of ASTROD missions \cite{ASTROD,Appourchaux2009}. ASTROD I mission concept is to use one spacecraft carrying a telescope, four lasers, two event timers and a clock with a Venus swing-by orbit. Two-way, two-wavelength laser pulse ranging will be used between the spacecraft in a solar orbit and deep space laser stations on Earth, to achieve the ASTROD I goals of testing GR with an improvement in sensitivity of 3 orders of magnitude and to measuring key solar system parameters with increased accuracy. Using the achieved accuracy of 3 ps in laser pulse timing and the demonstrated LISA (Laser Interferometer Space Antenna) Pathfinder drag-free capability, a simulation showed that accuracy of the determination of $\gamma$ will reach  $3\times 10^{-8}$.

The general concept of ASTROD is to have a constellation of drag-free spacecraft navigate through the solar system and range with one another using optical devices to map the solar-system gravitational field, to measure related solar-system parameters, to test relativistic gravity, to observe solar g-mode oscillations, and to detect gravitational waves. A baseline implementation of ASTROD, also called ASTROD, is to have two spacecrafts in separate solar orbits (one in inner solar orbit, the other in outer solar orbit), each carrying a payload of a proof mass, two telescopes, two 1-2 W lasers, a clock and a drag-free system, together with a similar spacecraft near Earth around one of the Lagrange points L1/L2. The three spacecrafts range coherently with one another using lasers to map solar-system gravity, to test relativistic gravity, to observe solar g-mode oscillations, and to detect gravitational waves. Since it will be after ASTROD I, we assume 1 ps timing accuracy and the drag-free performance of what LISA Pathfinder has achieved. With these requirements, the accuracy of the determination of $\gamma$ will reach $1\times 10^{-9}$ in 3.5 years \cite{Ni_2009}.

Super-ASTROD \cite{Ni_2009}, Odyssey \cite{Christophe2009}, SAGAS (Search for Anomalous Gravitation using Atomic Sensors) \cite{Wolf2009}, and OSS (Outer Solar System) \cite{Christophe2012} are four mission concepts to test fundamental physics and to explore the outer solar system. Their accuracy goals of measuring $\gamma$ are at $10^{-7}$-$10^{-8}$ level. All four mission concepts explore gravity at deep space to bridge the gap between inner solar-system tests and cosmological tests. They are most relevant to the detection/constraint of dark matter and dark energy, and to the tests of MOND models and the dark energy dynamical models. Ashby {\it et al.} \cite{Ashby2009}  proposed to use an optical clock and drag-free spacecraft to measure $\gamma$ about $10^{-8}$-level.

\begin{widetext}
	\begin{center}	
		\begin{table} 
			\caption{Ongoing/proposed experiments to measure $\gamma_W$}
			\label{table2}
			\begin{tabular}{|c|c|c|}		
				\hline
				% after \\: \hline or \cline{col1-col2} \cline{col3-col4} ...
				Ongoing/proposed experiment  & Aimed accuracy of $\gamma_W$& Type of experiment\\ \hline
				GAIA \cite{Gaia,Gaia2003,Gaia2016}& $1\times 10^{-5}$-$2\times 10^{-7}$ &  deflection \\
				Bepi-Colombo \cite{Bepi,PhysRevD.66.082001} & $2\times 10^{-6}$ &  retardation\\
				ASTROD I  \cite{ASTROD,Appourchaux2009} & $3\times 10^{-8}$ &  retardation \\
				ASTROD  \cite{ASTROD}  & $1\times 10^{-9}$ &   retardation \\
				Super-ASTROD  \cite{Ni_2009} & $1\times 10^{-8}$ &  retardation \\
				Odyssey \cite{Christophe2009} & $1\times 10^{-7}$ &  retardation \\
				SAGAS \cite{Wolf2009} & $1\times 10^{-7}$ &  retardation \\
				OSS  \cite{Christophe2012} & $1\times 10^{-7}$ &  retardation \\
				Ashby \cite{Ashby2009}  & $1\times 10^{-8}$ &  retardation \\
				\hline
			\end{tabular}
		\end{table}	
	\end{center}
	
\end{widetext} 

Ongoing/Proposed experiments are summarized in Table \ref{table2}. To measure $\gamma$  ($\gamma_W$ ) beyond the $10^{-6}$-$10^{-7}$ level, one would need to develop 2nd post-Newtonian solar-system ephemeris. Space gravitational missions will have high S/N ratio to test the propagation and polarization of GQFT \cite{Wu:2022aet,Wu:2022mzr}. In addition to test GQFT directly, the measurement of gravitomagnetic parameters in astrophysical phenomenon  (e. g., see \cite{Krishnan})  may give more possibilities to test GQFT together with other quantum gravity theories.

\begin{acknowledgments}
\noindent W.-T. Ni would like to thank Prof. M. S. Zhan, Prof. J. Wang and the Atomic Interferometry and Precision Measurement group for their great hospitality during the period from 2017 to 2024. This work was supported by the Technological Innovation 2030 "Quantum Communication and Quantum Computer" Major Project (Grants No. 2021ZD0300603 and No. 2021ZD0300604), and Spacetime precision measurement atomic interferometry research facility (Zhaoshan facility) pre-research and pre-manufacture (Grant No. S22H230102) 2023.12-2024.2.
\end{acknowledgments}

\bibliography{Wumodel2024}

\end{document}